\documentclass{article}
\usepackage{ismir,amsmath,cite,url}
\usepackage{graphicx}
\usepackage{color}
\usepackage{subcaption}
\usepackage{booktabs}       
\usepackage{float} 			
\usepackage{enumitem}
\usepackage[font={small}]{caption}
\title{Lyrics-Based Music Genre Classification Using a Hierarchical Attention Network}

\oneauthor
 {Alexandros Tsaptsinos} {ICME, Stanford University, USA\\ {\tt alextsap@stanford.edu}}

\begin{document}

\maketitle
\begin{abstract}
Music genre classification, especially using lyrics alone, remains a challenging topic in Music Information Retrieval. In this study we apply recurrent neural network models to classify a large dataset of intact song lyrics. As lyrics exhibit a hierarchical layer structure---in which words combine to form lines, lines form segments, and segments form a complete song---we adapt a hierarchical attention network (HAN) to exploit these layers and in addition learn the importance of the words, lines, and segments. We test the model over a 117-genre dataset and a reduced 20-genre dataset. Experimental results show that the HAN outperforms both non-neural models and simpler neural models, whilst also classifying over a higher number of genres than previous research. Through the learning process we can also visualise which words or lines in a song the model believes are important to classifying the genre. As a result the HAN provides insights, from a computational perspective, into lyrical structure and language features that differentiate musical genres.
\end{abstract}
\section{Introduction}\label{sec:introduction}
Automatic classification of music is an important and well-researched task in Music Information Retrieval (MIR)~\cite{mckinney2003feature}. Previous work on this topic has focused primarily on classifying mood~\cite{hu2010improving}, genre~\cite{mayer2008combination}, annotations~\cite{nam2012learning}, and artist~\cite{eghbal2015timbral}. Typically one or a combination of audio, lyrical, symbolic, and cultural data is used in machine learning algorithms for these tasks~\cite{mckay2010evaluating}.

Genre classification using lyrics presents itself as a natural language processing (NLP) problem. In NLP the aim is to assign meaning and labels to text; here this equates to a genre classification of the lyrical text. Traditional approaches in text classification have utilised $n$-gram models and algorithms such as Support Vector Machines (SVM), $k$-Nearest Neighbour (k-NN), and Na{\"i}ve Bayes (NB). 

In recent years the use of deep learning methods such as recurrent neural networks (RNNs) or convolutional neural networks (CNNs) has produced superior results and represent an exciting breakthrough in NLP~\cite{kalchbrenner2014convolutional, kim2014convolutional}. Whilst linear and kernel models rely on good hand-selected features, these deep learning architectures circumvent this by letting models learn important features themselves.

Deep learning has in recent years been utilised in several MIR research topics including live score following \cite{dorfer2016live}, music instrument recognition \cite{lostanlen2016deep}, and automatic tagging \cite{choi2016automatic}. In many cases, these approaches have led to significant improvements in performance. For example, Kum et al.~\cite{kum2016melody} utilise multi-column deep neural networks to extract melody on vocal segments while Southall et al.~\cite{southall2016automatic} approach automatic drum transcription using bidirectional recurrent neural networks.

Neural methods have further been utilised for the genre classification task on audio and symbolic data. Sigtia and Dixon \cite{sigtia2014improved} use the hidden states of a neural network as features for song on which a Random Forest classifier was built, reporting an accuracy of 83\% among 10 genres. Costa et al.~\cite{costa2017evaluation} compare the performance of CNNs in genre classification through spectrograms with respect to results obtained through hand-selected features and SVMs. Jeong and Lee \cite{jeong2016learning} learn temporal features in audio using a deep neural network and apply this to genre classification. However, not much research has looked into the performance of these deep learning methods with respect to the genre classification task on lyrics. Here, we attempt to remedy this situation by extending deep learning approaches to text classification to the particular case of lyrics. 

Hierarchical methods attempt to use some sort of structure of the data to improve the models and have previously been utilised in vision classification tasks~\cite{seo2016progressive}. Yang et al.~\cite{yang2016hierarchical} propose a hierarchical attention network (HAN) for the task of document classification. Since documents often contain structure whereby words form to create sentences, sentences to paragraphs, etc. they introduce this knowledge to the model, resulting in superior classification results. It is evident that songs and, in particular, lyrics similarly contain a hierarchical composition: Words combine to form lines, lines combine to form segments, and segments combine to form the whole song. A segment of a song is a verse, chorus, bridge, etc.~of a song and typically comprises several lines. The hierarchical nature of songs has been previously exploited in genre classification tasks with Du et al.~\cite{du2016new} utilising hierarchical analysis of spectrograms to help classify genre. 

Here, we propose application of an HAN for genre classification of intact lyrics. We train such a network, allowing it to apply attention to words, lines, and segments. Results show the network produces higher accuracies in the lyrical classification task than previous research and from the attention learned by the network we can observe which words are indicative of different genres.

The remainder of the paper is structured as follows. In \secref{sec:methods} we describe our methods, including the dataset and a description of the HAN. In \secref{sec:experiments} we provide results and visualisations from our experiments. We conclude with a discussion in \secref{sec:discussion}.

\section{Methods}\label{sec:methods}

\subsection{Dataset}\label{subsec:dataset}
Research involving song lyrics has historically suffered from copyright issues. Consequently most previous literature has utilised count-based bag-of-words lyrics. In this format, structure and word order are lost, and it has been shown that utilising intact lyrics reveals superior results in classification tasks~\cite{fell2014lyrics,smith2012your}.

Seeking an intact lyrics corpus for the present study, we obtained a collection of lyrics through a signed research agreement with LyricFind\footnote{\url{http://lyricfind.com/}}. This corpus has been used in the past to study novelty~\cite{ellis2015quantifying} and influence~\cite{atherton2016said} in lyrics. The complete set contained 1,039,151 song lyrics in JSON format, as well as basic metadata including artist(s) and track name. As the corpus provided no genre information, we aggregated it ourselves using the iTunes Search API\footnote{\url{http://apple.co/1qHOryr}}, extracting the value for the \texttt{primaryGenreName} key as baseline truth. Several different sources were not used for consistency reasons with iTunes found to be the largest, easily accessible source with reasonable genre tags. This unfortunately still greatly reduced the size of the dataset due to the sparse iTunes database. We then further removed any songs that were linked with a genre tag of `Music Video', leaving a dataset comprising 244 genres. As this dataset had a very long tail of sparse genres, we further filter the dataset via two methods. Firstly we remove any genres with less than 50 instances, giving a dataset of size 495,188 lyrics and 117 genres. Secondly we retain only the top 20 genres, giving a dataset of 449,458 lyrics. We note also that the dataset originally contained various versions of the same lyrics, due to the prevalence of cover songs; we retain only one of these versions chosen at random. The song lyrics are split into lines and segments which we tokenised using the \texttt{nltk} package\footnote{\url{http://www.nltk.org/}} in Python. We split the dataset into a rough split of 80\% for training, 10\% for validation, and 10\% for testing. All pre-processing was done via Python with the neural networks built using Tensorflow\footnote{\url{https://www.tensorflow.org/}}.

\subsection{Hierarchical Attention Networks}\label{sec:HANs}
\begin{figure}[t]
  \centering
  \includegraphics[width=\columnwidth]{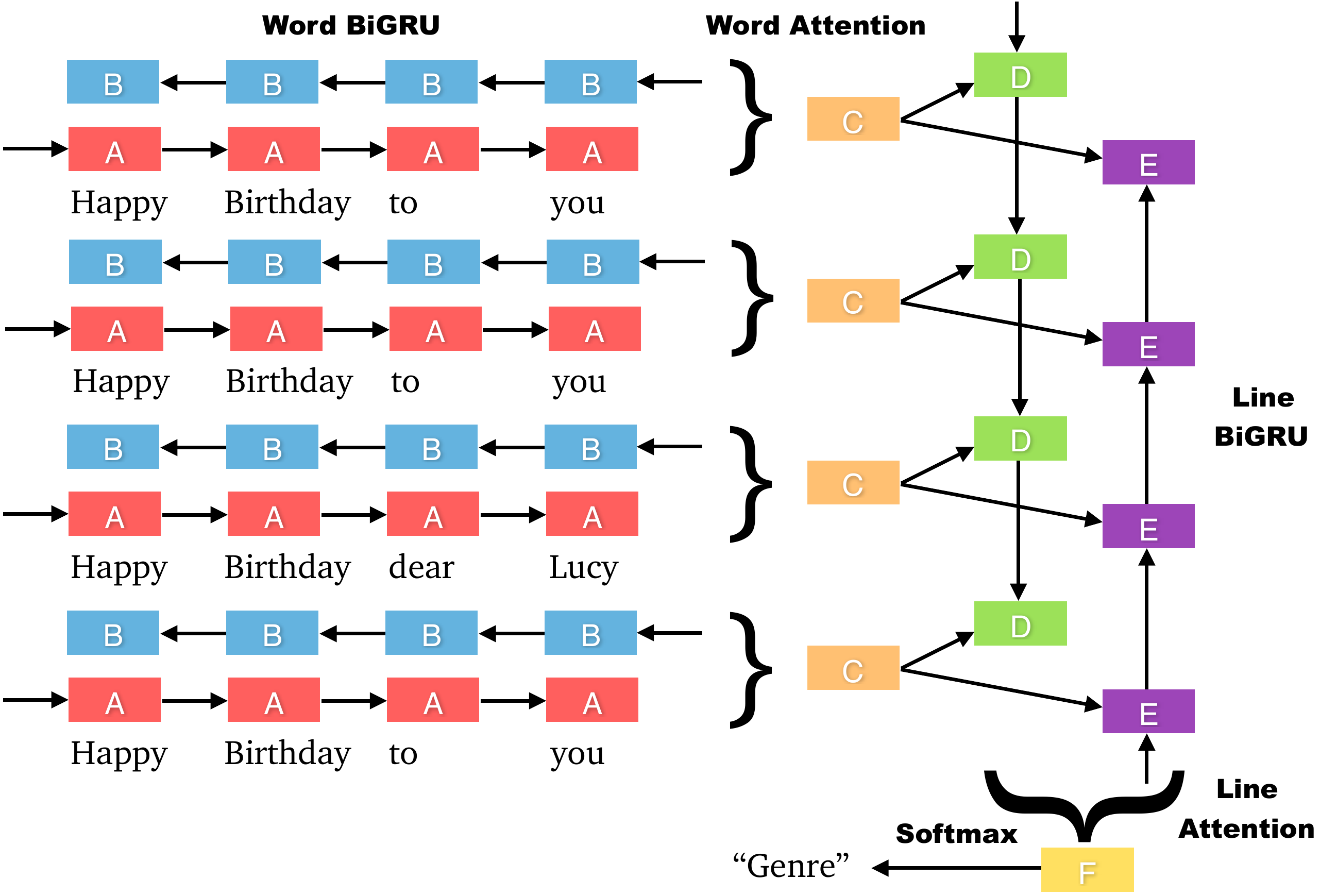}
  \caption{Representation of the HAN architecture; boxes represent vectors. A and B vectors represent the hidden states for the forward and backward pass of the GRU at the word level, respectively. The line vectors C are then obtained from these hidden states via the attention mechanism. The D and E vectors represent the forward and backward pass of the GRU at the line level, respectively. The song vector F is then obtained from these hidden states via the attention mechanism. Finally classification is performed via the softmax activation function.}
  \label{fig:architecture}
\end{figure}

The structure of the model follows that of Yang et al.~\cite{yang2016hierarchical}. Each layer is run through a bidirectional gated recurrent unit (GRU) with attention applied to the output. The attention weights are used to create a vector via a weighted sum which is then passed as the input to the next layer. A representation of the architecture for the example song of `Happy Birthday' can be seen in \figref{fig:architecture}, where the layers are applied at the word, line, and song level. We briefly step through the various components of the model.

\subsubsection{Word Embeddings}
An important idea in NLP is the use of dense vectors to represent words. A successful methodology proposes that similar words have similar context and thus vectors can be learned through their context, such as in the word2vec model~\cite{mikolov2013distributed}. Pennington et al.~\cite{pennington2014glove} propose the GloVe method which combines global matrix factorisation and local context window methods to produce word vectors that outperform previous word2vec and SVM based models.

Here we take as our vocabulary the top 30,000 most frequent words from the whole LyricFind corpus, including those from songs we did not match with a genre. We train 100-dimensional GloVe embeddings for these words using methods obtained from the GloVe website\footnote{\url{http://nlp.stanford.edu/projects/glove/}}. Previous research has shown that retraining these word vectors over the extrinsic task at hand can improve results if the dataset is large enough~\cite{collobert2011natural}. In a preliminary genre classification task we found that retraining these word embeddings did improve accuracy, and so we let our model learn superior embeddings to those provided by GloVe \cite{pennington2014glove}.

\subsubsection{Gated Recurrent Units}\label{subsubsec:GRUs}
Introduced by Chung et al.~\cite{chung2014empirical}, GRUs are a form of gating mechanism in RNNs designed to help overcome the struggle to capture long-term dependencies in RNNs. This is achieved by the introduction of intermediate states between the hidden states in the RNN. An update gate $z_{t}$ is introduced to help determine how important the previous hidden state is to the next hidden state. A reset gate $r_{t}$ is introduced to help determine how important the previous hidden state is in the creation of the next memory. The hidden state is $h_{t}$, whilst new memory is computed and stored in $\tilde{h}_{t}$. Mathematically we describe the process as
\begin{align}
z_{t} &= \textrm{sigmoid}\left(W_{z} x_{t} + U_{z} h_{t-1} + b_{z} \right) \\
r_{t} &= \textrm{sigmoid}\left(W_{r} x_{t} + U_{r} h_{t-1} + b_{r} \right) \\
\tilde{h}_{t} &= \textrm{tanh}\left(W_{h} x_{t} + r_{t} \circ U_{h} h_{t-1}+ b_{h} \right) \\
h_{t} &= (1-z_{t}) \circ h_{t-1} + z_{t} \circ \tilde{h}_{t} ,
\end{align}
where $x_{t}$ is the word vector input at time-step $t$, $\circ$ is the Hadamard product, and $\textrm{sigmoid}$ is the sigmoid activation function. $W_{z}$, $U_{z}$, $W_{r}$, $U_{r}$, $W_{h}$, and $U_{h}$ are weight matrices randomly initialised and to be learned by the model along with the $b_{z}$, $b_{r}$, and $b_{h}$ bias terms. Bias terms were not included in the original model by Chung et al.~\cite{chung2014empirical}, however have been included here as in Jozefowicz et al.~\cite{jozefowicz2015empirical}. 

\subsubsection{Hierarchical Attention}\label{subsubsec:HA}
Attention was first proposed by Bahdanau et al.~\cite{bahdanau2014neural} with respect to neural machine translation to allow the model to learn which words were more important in the translation objective. Along the lines of that study, we would like our model to learn which words are important in classifying genre and then apply more weight to these words. Similarly, we can apply attention again on lines or segments to let the model learn which lines or segments are more important in classification.

Given input vectors $h_{i}$ for $i = 1, \dots, n$ the attention mechanism can be formulated as
\begin{align}
u_{i} &= \textrm{tanh}\left(W_{a} h_{i} + b_{a} \right) \\
\alpha_{i} &= \frac{\exp(u_{i}^{T} u_{a})}{\sum_{k=1}^{n} \exp(u_{k}^{T} u_{a})} \\
s &= \sum_{i=1}^{n} \alpha_{i} h_{i},
\end{align}
where $s$ is the output vector passed to the next layer consisting of the weighted sum of the current layers vectors. Parameters $W_{a}$, $b_{a}$, and $u_{a}$ are learned by the model after random initialisation.

One layer of the network takes in vectors $x_{1}, \dots, x_{n}$, applies a bidirectional GRU to find a forward hidden state $\overrightarrow{h}_{j}$ and a backward hidden state $\overleftarrow{h}_{j}$, and then uses the attention mechanism to form a weighted sum of these hidden states to output as the representation. Letting $GRU$ indicate the output of a GRU and $ATT$ represent the output from an attention mechanism, one layer is formulated as
\begin{align}
\overrightarrow{h}_{j} &= \overrightarrow{GRU}(x_{j}), \label{eq:layer1} \\
\overleftarrow{h}_{j} &= \overleftarrow{GRU}(x_{j}), \\
h_{j} &= [\overrightarrow{h}_{j} ; \overleftarrow{h}_{j}], \\
s &= ATT(h_{1}, \dots, h_{L}).
\label{eq:layer4}
\end{align}
Our HAN consists of two layers, one at the word level, and one at the line/segment level. Consider a song of $L$ lines or segments $s_{j}$, each consisting of $n_{j}$ words $w_{ij}$. Let $E$ be the pre-trained word embedding matrix.  Letting $LAY$ represent the dimension reduction operation of a layer in the network as in Eqns~\ref{eq:layer1}--\ref{eq:layer4} the whole HAN can be formulated for $i = 1, \dots, n_{j}$ and $j = 1, \dots, L$ as
\begin{align}
x_{ij} &= Ew_{ij} \\
s_{j} &= LAY(x_{1j}, \dots, x_{n_{j}j}), \\
s &= LAY(s_{1}, \dots, s_{L}).
\end{align}
Each layer has its own set of GRU weight matrix and bias terms to learn, as well as its own attention weight matrix, bias terms, and relevance vector to learn. 

\subsubsection{Classification}\label{subsubsec:classification}
With the song vector $s$ now obtained, classification is performed by using a final softmax layer
\begin{align}
p = \text{softmax}\left(W_{p} s + b_{p}\right),
\end{align}
where intuitively we take the entry of highest magnitude as the prediction for that song.

To train the model we minimise cross-entropy loss over $K$ songs
\begin{align}
J = - \sum_{k=1}^{K} \log (p_{d_{k}k}),
\end{align}
where $d_{k}$ is the true genre label for that song.

\section{Experiments}\label{sec:experiments}

\subsection{Baseline Models}\label{subsec:baselineclass}
We compare the performance of the HAN against various baseline models. 
\begin{enumerate}[leftmargin=*]
\setlength\itemsep{0em}
\item Majority classifier (MC): `Rock' is the most common genre in our dataset. The MC simply predicts `Rock'.
\item Logistic regression (LR): A LR run on the average song word vector produced from the GloVe embeddings.
\item Long Short-Term Memory (LSTM): An LSTM, treating the whole song as a single sequence of words and use max-pooling of the hidden states for classification. Fifty hidden units were used in the LSTM and each song had a maximum of 600 words. For full discussion of the LSTM framework see Hochreiter and Schmidhuber~\cite{hochreiter1997long}.
\item Hierarchical network (HN-L): The HN structure in the absence of attention run at the line level. At each layer all of the representations are simply averaged to produce the next layer input.
\end{enumerate}

For LR, LSTM, and HN-L we let the model retrain the word embeddings as it trained.

\subsection{Model Configuration}\label{subsec:config}
The lyrics are padded/truncated to have uniform length. In the line model, each line has a maximum of 10 words and a maximum of 60 lines. In the segment model each segment has a maximum of 60 words and a maximum of 10 segments. Fifty hidden units are utilised in the bidirectional GRUs, whilst one hundred states are output from the attention mechanisms. Before testing the model, hyper-parameters were tuned on the validation set. Dropout \cite{srivastava2014dropout} and gradient clipping \cite{pascanu2013difficulty} were both found to benefit the model. We dropout at each layer with probability $p=0.5$ and gradients are clipped at a maximum norm of 1 in the backpropogation. We utilise a mini-batch size of 64 and optimise using RMSprop~\cite{tieleman2012lecture} with a learning rate of $0.01$. The models were all run until their validation loss did not decrease for 3 successive epochs. In all the HAN models, this occurred between the 5th and 8th epoch.

The code to train the model and perform the experiments described are made publicly available\footnote{\url{https://github.com/alexTsaptsinos/lyricsHAN}}.

\subsection{Results}\label{subsec:results}
For both dataset sizes we run the baseline models and the HAN at the line and segment level. Let HAN-L represent running over lines and HAN-S represent running over segments. The test accuracies are seen in \tabref{tab:accs}.
\begin{table}
  \centering
  \begin{tabular}{l c c}
    \toprule
    Model & 117 Genres & 20 Genres   \\
    \midrule
    \textbf{MC} & 24.71 & 27.17 \\
    \textbf{LR} & 35.21 & 38.13 \\
    \textbf{LSTM} & 43.66 & 49.77 \\
    \textbf{HN-L} & 45.85 & 49.09 \\
    \textbf{HAN-L} & 46.42 & 49.50  \\
    \textbf{HAN-S} & 45.05 & 47.60 \\
     \bottomrule
  \end{tabular}
  \caption{Genre classification test accuracies for the two datasets (\%) using majority classifier (MC), logistic regression (LR), Long Short-Term Model (LSTM), hierarchical network (HN-L), and line- and segment-level HAN (HAN-L, HAN-S).}
  \label{tab:accs}
\end{table}

From the results we see a trend between model complexity and classification accuracy. The very simple majority classifier performs weakest and is improved upon by the simple logistic regression on average bag-of-words. The neural-based models perform better than both of the simple models. The LSTM model, which takes into account word order and tries to implement a memory of these words, gives performances of 43.66\% and 49.77\%, outperforming the HAN on the 20-genre dataset. Over the 117-genre dataset the best performing models were the HANs, with a highest accuracy of 46.42\% when run over lines. It is observed that for the simpler 20-genre case, the more complex HAN is not required since the simpler LSTM beats it, although the LSTM took almost twice as long to train as the HAN. However for the more challenging 117-genre case, the HAN-L outperforms the LSTM, perhaps picking up on more of the intricacies of rarer genres.

In both cases the HAN run at the line level produced superior results than that run over the segment level, giving a bump of roughly 1.4\% and 1.9\% in the 117-genre and 20-genre datasets, respectively. The HN-L, which is run at the line level, additionally outperforms the HAN at segment level. This indicates that the model performs better when looking at songs line by line rather than segment by segment. In the HAN-L the model can pick up on many repeated lines or lines of a similar ilk, rather than the few similar segments it attains in the HAN-S, and this may be attributive to the better performance. The network does benefit from the inclusion of attention, with HAN-L classifying with higher accuracies than HN-L. This increase is marginal and requires an increased cost, however allows for the extraction of attention in the visualisations of the following section. 

As expected, classifying over the 20-genre dataset has given boosts of roughly 3\% and 2.5\% in the HAN-L and HAN-S, respectively. It is interesting to note that discarding roughly 10\% of the data by only keeping roughly a sixth of the genres has not strengthened the model by much. Given the similarity of recognition performance between the two datasets, even with the simplest of models, it is likely that the extra genres are predominantly noise added to the 20-genre dataset. With the HAN-L outperforming the LSTM over the 117-genre dataset this then indicates that the model is more robust to noise.

\begin{figure}[t!]
\centering
\includegraphics[width=0.87\columnwidth]{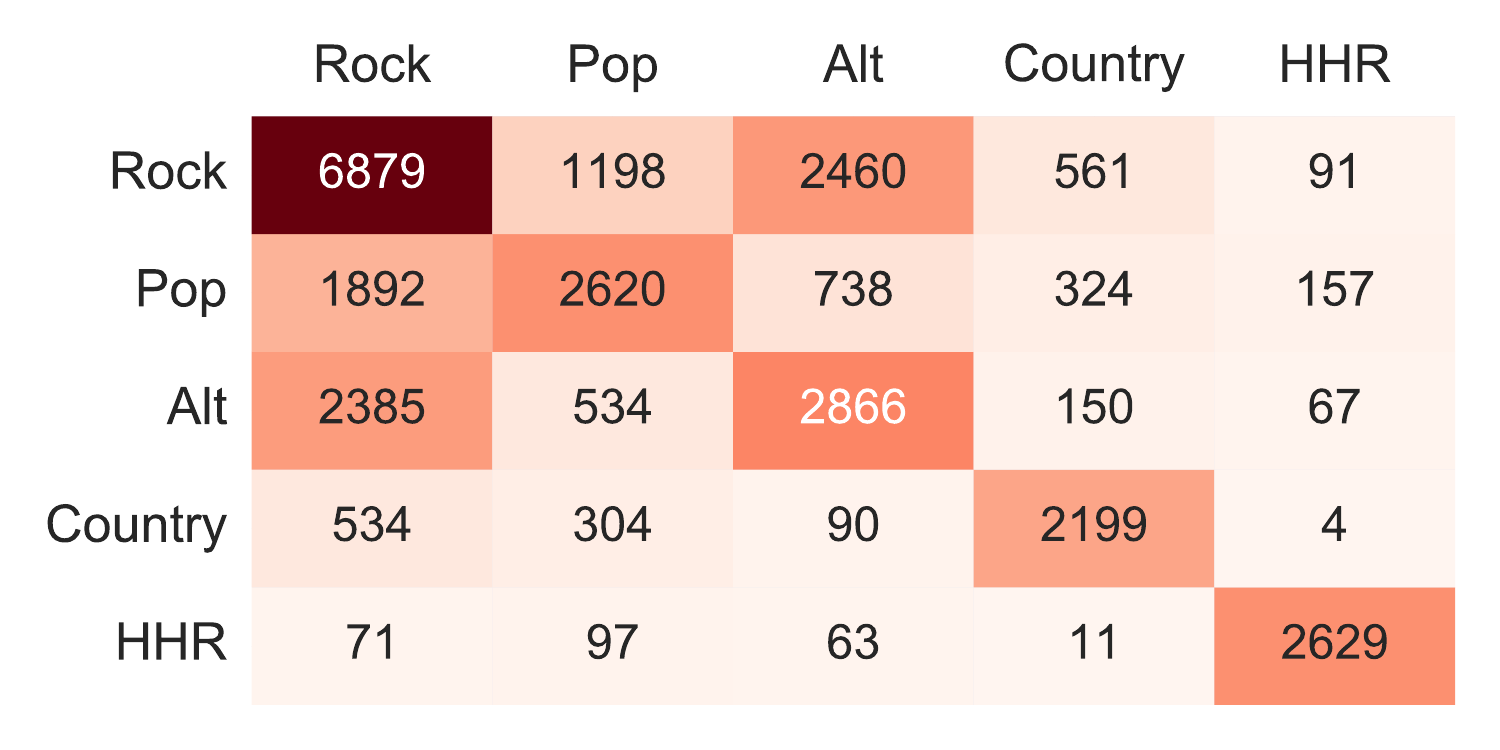}
\caption{HAN-L confusion matrix for Rock, Pop, Alternative (Alt), Country, and Hip-Hop/Rap (HHR) genres over larger (117-genre) dataset. Rows represent true genre, whilst columns are predicted.}
\label{fig:confusion}
\end{figure}

The confusion matrix for HAN-L run over the larger dataset for the top 5 genres can be seen in \figref{fig:confusion}. We can see from the matrix that Rock, Pop, and Alternative (Alt) are all commonly confused; the model predicts Rock for Alternative almost as many times as it does Alternative. As the most common genre in the dataset by about 30,000 it is unsurprising to see the model try and predict Rock more often, and it is unclear whether a person would be able to distinguish between the lyrics of these genres. However, we see that both Country and Hip-Hop/Rap (HHR) are more separated. With their distinct lyrical qualities, especially in the case of Hip-Hop/Rap, this is an encouraging result indicating that the model has learned some of the qualities of both these genres. 

\begin{figure*}[t!]
    \centering
    \begin{subfigure}{\textwidth}
        \includegraphics[scale=0.6, trim={3cm 0 0 0}]{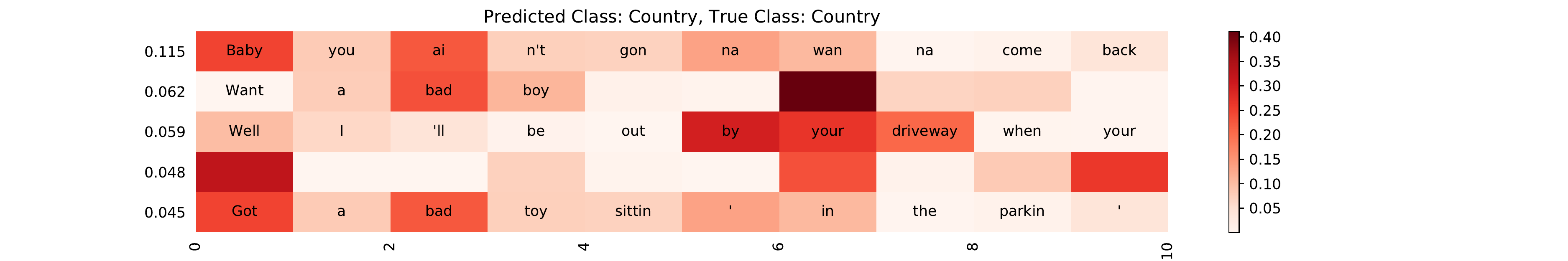}
    \end{subfigure}
    ~ 
    \begin{subfigure}{\textwidth}
        \includegraphics[scale=0.6, trim={3cm 0 0 0}]{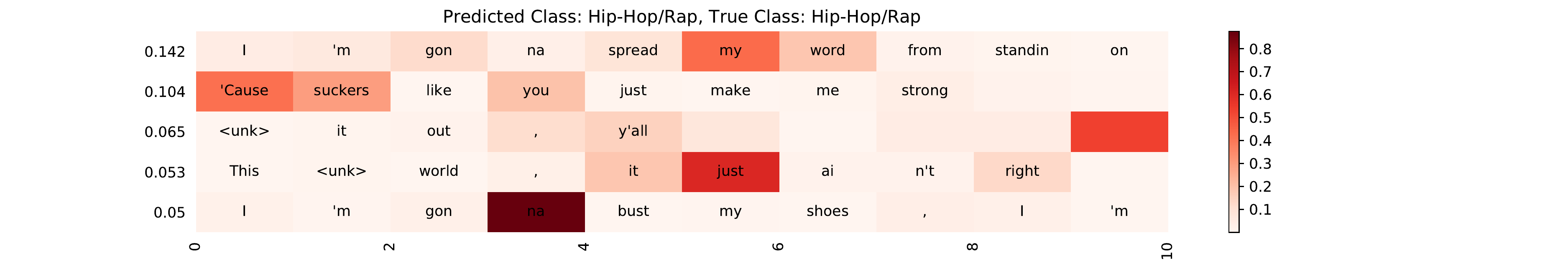}
    \end{subfigure}
    ~
    \begin{subfigure}{\textwidth}
        \includegraphics[scale=0.6, trim={3cm 0 0 0}]{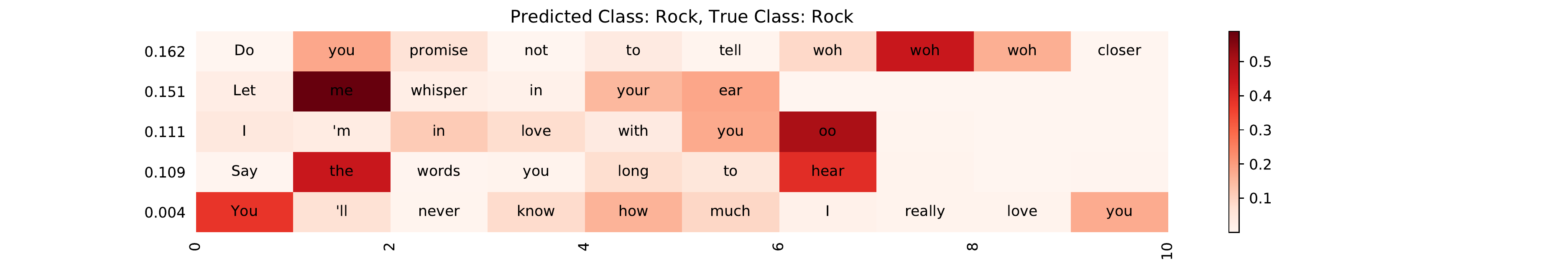}
    \end{subfigure}
    \caption{Weights applied by the HAN-L for song lyrics that were correctly classified. Line weights appear to the left of each line and word weights are coloured according to the respective colorbars on the right.}
    \label{fig:weights_correct}
\end{figure*}

\subsubsection{Attention Visualisation}\label{subsubsec:attentionviz}
To help illustrate the attention mechanism, we feed song lyrics into the HAN-L and observe the weights it applies to words and lines. For each song we extract the 5 most heavily weighted lines and a visualisation of their weights and the individual word weights for a few different correctly predicted song lyrics can be seen in Figure \ref{fig:weights_correct}.

From these visualisations we notice that the model has placed greater weights on words we may associate with a certain genre. For example `baby' and `ai' are weighted heavily in the Country song, and the most heavily weighted line in that song is characteristically Country. The model has placed great weight on a blank line, indicating the break between segments; it is unclear whether the model is learning to place importance on how songs are segmented and the number of segments occurring. In the Hip-Hop/Rap song the model places attention on colloquially spelled words `cause' and `gonna'. Although not included here, it was observed that for many rap songs swear words and racial terms were heavily weighted. The model picks up the `woh' and `oo' in the Rock song and also heavily weights occurrences of second-person determiner `your' and pronoun `you'. It was found that for many Rock songs this was the case.

\begin{figure*}[t!]
    \centering
    \begin{subfigure}{\textwidth}
        \includegraphics[scale=0.6, trim={3cm 0 0 0}]{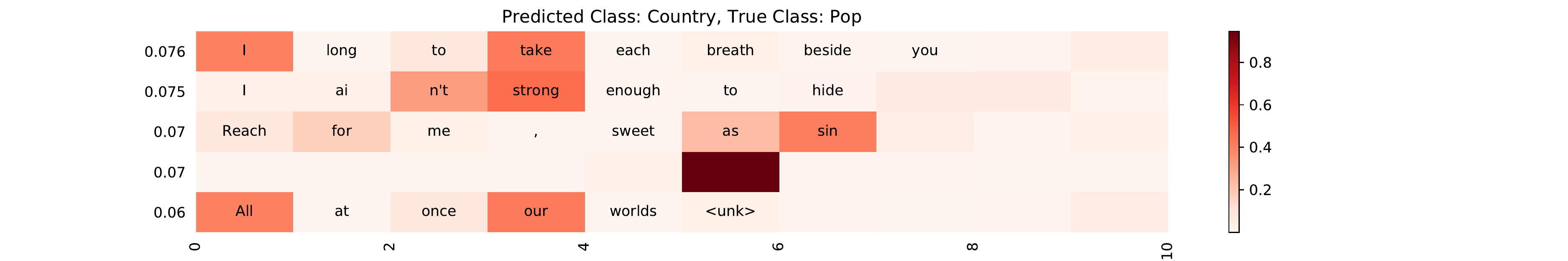}
    \end{subfigure}
    ~ 
    \begin{subfigure}{\textwidth}
        \includegraphics[scale=0.6, trim={3cm 0 0 0}]{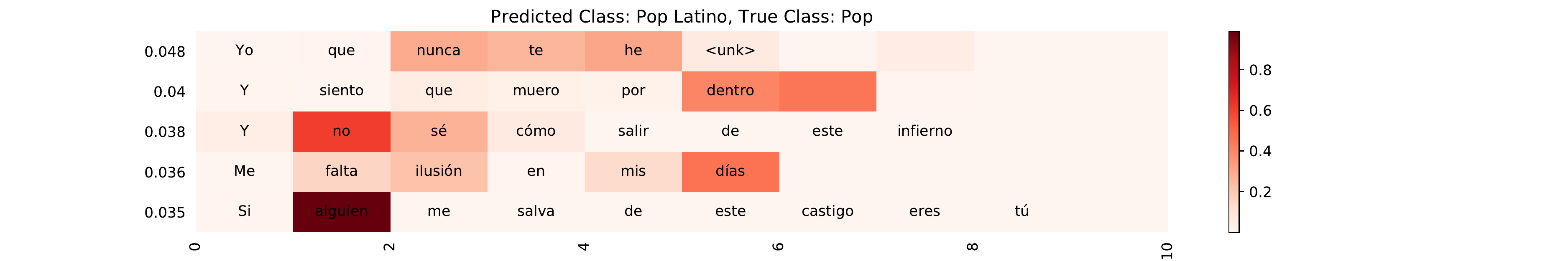}
    \end{subfigure}
    ~
    \begin{subfigure}{\textwidth}
        \includegraphics[scale=0.6, trim={3cm 0 0 0}]{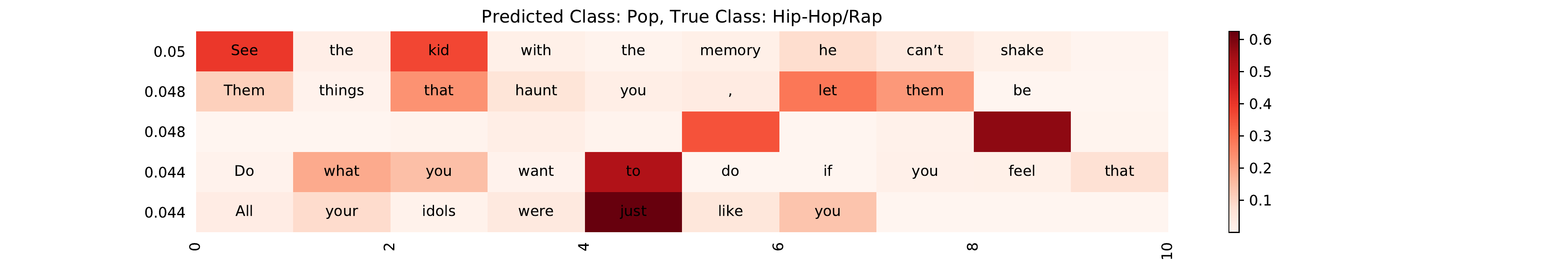}
    \end{subfigure}
    \caption{Weights applied by the HAN-L for song lyrics that were incorrectly classified. Line weights appear to the left of each line and word weights are coloured according to the respective colorbars on the right.}
    \label{fig:weights_wrong}
\end{figure*}

In addition some visualisations of lyrics that were incorrectly classified by the HAN-L can be seen in Figure \ref{fig:weights_wrong}. We observe the model predicting Country for a Pop song, applying weights to `sin' and `strong' which could be characteristic of Country songs. The dataset contains songs with foreign language lyrics. Here we observe a song with Spanish lyrics classed as Pop Latino by the model whilst iTunes deems it Pop. This seems like a fair mistake for the model to have made since it has evidently recognised the Spanish language. The model also incorrectly classifies the Hip-Hop/Rap song as Pop. In the 5 most heavily weighted lines we do not spot any instances of language that indicate a Hip-Hop/Rap song and we hypothesise that the genericness of the lyrics has led the model to predict Pop.

\section{Discussion}\label{sec:discussion}
Genre is an inherently ambiguous construct, but one that plays a major role in categorising musical works~\cite{mckay2006musical,sordo2008quest}. From one standpoint, genre classification by lyrics will always be inherently flawed by vague genre boundaries and many genres borrowing lyrics and styles from one another. Previous research has shown that lyrical data performs weakest in genre classification compared to other forms of data~\cite{mckay2010evaluating}. As a consequence, this problem is not as well researched and preference has been given to other methods.

SVMs, k-NN, and NB have been heavily used in previous lyrical classification research. In addition very rarely has research looked into classifying more than between 10 genres despite the prevalence of clearly many more genres. Fell and Sporleder classify among 8 genres using $n$-grams along with other hand-selected features to help represent vocabulary, style, structure, and semantics~\cite{fell2014lyrics}. Ying et al.\ make use of POS tags and classify among 10 genres using SVMs, $k$-NN, NB with a highest accuracy of 39.94\%~\cite{ying2012genre}. McKay et al.\ utilise hand-selected features to produce classification accuracies of 69\% among 5 genres and 43\% among 10 genres~\cite{mckay2010evaluating}.

In this paper we have shown that an HAN and other neural-based methods can improve on the genre classification accuracy. In large part this model has beaten all previously reported lyrical-only genre classification model accuracies, except for the classification among 5 genres. Whilst having been trained on different datasets the jump in classification accuracies achieved by the HAN and LSTM across the 20-genre datasets compared to previous research indicate that neural structures are clearly beneficial. However, with very similar results between the neural structures it is still unclear what the optimal neural structure may be and there is certainly room for further experimentation. We have shown that the HAN works better with layers at the word, line, and song level rather than word, segment, and song level. One known issue of the present dataset is that iTunes attributes genres by artist, not by track; this is a problem for artists whose work may cover multiple genres and is something that should be addressed in the future. A larger issue concerns the accuracy of the iTunes genre labels more generally, especially for the larger 117-genre dataset which naturally includes more subjective and vague genre definitions.

Visualisations of the weights the HAN applies to words and lines were produced to help see what the model was learning. In a good amount of cases, words and lines were heavily weighted that were cohesive with the song genre; however, this was not always the case. We note that in general the model tended to let one word dominate a single line with the greatest weight. However this was not as apparent across lines, with weights among lines more evenly spread. With a large amount of foreign-language lyrics also present in the dataset, an idea for further research is to build a classifier that identifies language, and from there classifies by genre. Any such research would be inhibited, however, by the lack of such a rich dataset to train on.

To produce a state-of-the-art classifier it is evident that the classifier must take into account more than just the lyrical content of the song. Mayer et al.\ combine audio and lyrical data to produce a highest accuracy of 63.50\% within 10 genres via SVMs~\cite{mayer2008combination}. Mayer and Rauber then use a cartesian ensemble of lyric and audio features to gain a highest accuracy of 74.08\% within 10 genres~\cite{mayer2011musical}. Further research could look into employing this hierarchical attention model to the audio and symbolic data, and combining with the lyrics to build a stronger classifier. Employment of the HAN in the task of mood classification via sentiment analysis is another possible area of research. In addition the HAN could be extended to include both a layer at the line and segment level, or even at the character level, to explore performance.

\section{Acknowledgements}
Many thanks to Will Mills and Mohamed Moutadayne from LyricFind for providing access to the data, and the ISMIR reviewers for their helpful comments.

\bibliography{writeup}

\end{document}